\documentclass{Interspeech}

\interspeechcameraready

\title{Layer-wise Investigation of Large-Scale Self-Supervised Music Representation Models
}
\author[affiliation={1}]{Yizhi}{Zhou}
\author[affiliation={2}]{Haina}{Zhou}
\author[affiliation={3}]{Hangting}{Chen \texorpdfstring{$\dagger$}{$\dagger$}}

\affiliation{}{NanJing University}{China}
\affiliation{}{Shanghai Jiao Tong University}{China}
\affiliation{}{Tencent AI Lab }{China}
\email{zhouyz@lamda.nju.edu.cn, hainazhu@sjtu.edu.cn, erichtchen@tencent.com}
\keywords{Self-supervised learning, representation analysis, music information retrieval }

\usepackage{comment}
\usepackage{cite}
\usepackage{amsmath,amssymb,amsfonts}
\usepackage{algorithmic}
\usepackage{graphicx}
\usepackage{textcomp}
\usepackage{xcolor}
\usepackage{float}
\usepackage{subcaption}
\usepackage{enumitem}
\usepackage{lipsum}

\begin{document}

\maketitle
\newcommand\blfootnote[1]{%
\begingroup
\renewcommand\thefootnote{}\footnote{#1}%
\addtocounter{footnote}{-1}%
\endgroup
}

\begin{abstract}
    
    Recently, pre-trained models for music information retrieval based on self-supervised learning (SSL) are becoming popular, showing success in various downstream tasks. However, there is limited research on the specific meanings of the encoded information and their applicability. Exploring these aspects can help us better understand their capabilities and limitations, leading to more effective use in downstream tasks.

In this study, we analyze the advanced music representation model MusicFM and the newly emerged SSL model MuQ. We focus on three main aspects: (i) validating the advantages of SSL models across multiple downstream tasks, (ii) exploring the specialization of layer-wise information for different tasks, and (iii) comparing performance differences when selecting specific layers. Through this analysis, we reveal insights into the structure and potential applications of SSL models in music information retrieval.
\end{abstract}

\blfootnote{
Work performed during an internship at Tencent AI Lab.

$\dagger$ Corresponding Author.}

\section{Introduction}
Self-superviesed learning (SSL) has been used to train large fundamental models, including audio signal processing models \cite{li2023mert, won2023musicfm}. One of the most significant advantages of SSL models is their ability to leverage unlabeled audio data, enabling the possibility of training with large-scale datasets. Recently, numerous SSL models and methods leveraging SSL models for downstream tasks have been proposed, including those in the field of music information retrieval \cite{Spijkervet_Burgoyne_2021, Castellon_Donahue_Liang_2021, mccallum2022supervised}. However, the features extracted by these models and the information contained within them have not been thoroughly explored, which may increase the difficulty and error rate in developing downstream tasks using these features. 

We aim to enhance our understanding of the information encoded in pre-trained models by investigating how features evolve across layers, and how these features can be better utilized.  In this work, we analyze the recent state-of-the-art music SSL model, MusicFM, and the newly emerged MuQ model. Our main findings are shown as follows:

\begin{enumerate}
    \item SSL models in Music Information Retrieval (MIR) tasks was demonstrated effective through comprehensive experiments.
    

    \item Feature representations progressively evolve from Acoustic-level to Semantic-level in the model.
\end{enumerate}

\section{Related work}
\label{sec:format}

Music Information Retrieval (MIR) tasks primarily aims to extract useful information from raw audio signals, including Genre classification, Key detection, and so on. However, research in this area has been hindered by the limitations in data access \cite{Chen_Keast_Moody_Moriarty_Villalobos_Winter_Zhang_Lyu_Freeman_Wang_etal._2019}. To address this challenge, self-supervised methods \cite{Castellon_Donahue_Liang_2021} have been introduced to the MIR field due to their ability to utilize unlabeled data and establish a unified upstream and downstream architecture \cite{mccallum2022supervised}, including wav2vec \cite{Schneider_Baevski_Collobert_Auli_2019_wav2vec, Baevski_Zhou_Mohamed_Auli_Ai_w2v2} that based on Contrastive Learning,  MERT\cite{li2023mert} that adapted Hu-BERT \cite{Hsu_Bolte_Tsai_Lakhotia_Salakhutdinov_Mohamed_2021_hubert} to conduct a framework  for music representation. More recently, researchers introduced  MusicFM \cite{won2023musicfm} and MuQ\cite{zhu2025muq} base on BEST-RQ \cite{chiu2022bestrq}, proving its generalization ability on sequence-level and frame-level MIR tasks.

Some research has been dedicated to analyzing the reasons behind the high performance of speech SSL models \cite{Chen_Wu_Wang_Liu_Chen_Wang_Liu_Li_Wu_Yu_etal._2022}, and there are studies that have conducted layer-wise analysis on speech SSL models of feature information changes and the performance differences in downstream speech tasks \cite{Pasad_Chou_Livescu_2021}. Additionally, recent studies have comprehensively evaluated the performance of various mainstream SSL models on speech tasks \cite{yang2024largescaleevaluationspeechfoundation}. We utilize MARBLE benchmark\cite{yuan2023marble} to analyze music SSL models, including comprehensive and layer-wise analysis. To  our knowledge, this is the first work to analyze music SSL models on the layer-wise granularity.



\section{Analysis Methods and Details}

\begin{figure}[t]
    \centering
    \includegraphics[width=0.9\linewidth]{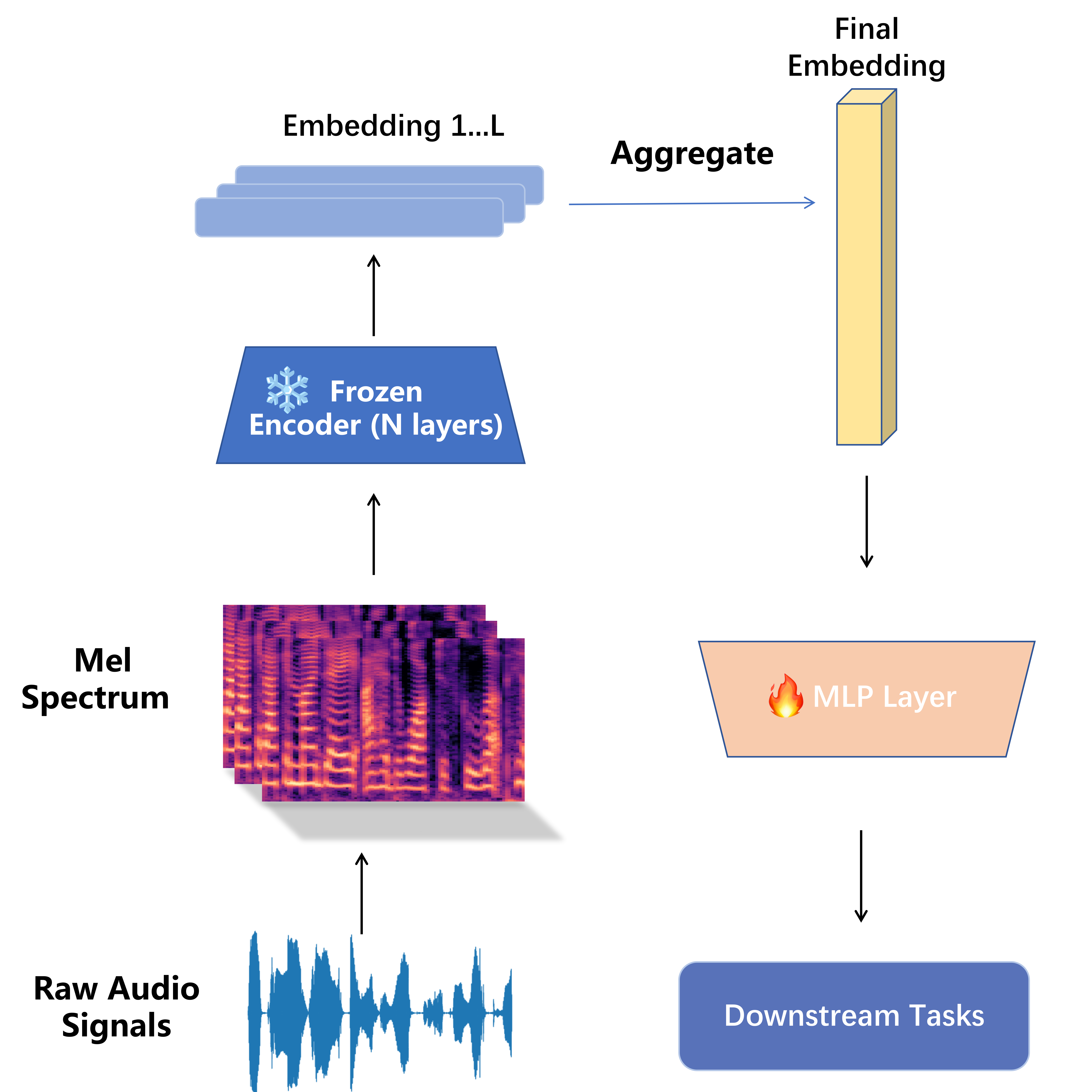}
    \caption{Model Framework, illustrates the paradigm of using SSL models for downstream tasks.}
    \label{fig:model-pipe}
\end{figure}

\subsection{Foundation Model} The foundation models are pre-trained on a large amount of unlabeled data to learn high-dimensional representations of audio. It aims to compress and denoise long raw audio signals, retaining as much information in signals as possible. The training objective is to predict tokens that have been randomly masked within a sequence. similar to BERT \cite{Devlin_Chang_Lee_Toutanova_2019_bert}. In audio processing, authors used a variety of feature extractors to generate the tokens for training. 
MusicFM  use random projection quantizer as feature extractor following the settings in BEST-RQ .



To further explore the shared characteristics of the SSL models, we also used a SSL model called MuQ with the same structure\cite{zhu2025muq}. The training labels for this model were tokens extracted from Mel Residual Vector Quantization (Mel-RVQ), which quantizes the Mel spectrum using a linear RVQ \cite{barnes1996rvq}. 

After the pre-training, the models are considered capable of extracting features from previously unseen music audio. During evaluating, the Mel spectrum is extracted from raw audios, resampled, then fed into multiple layers of encoders for context learning as in \eqref{eq:1},\eqref{eq:2}. At each layer $i$ from 1 to $L$ ($L$ is the number of encoder layers), we can obtain sequence embeddings $z_i$ for audio signals. These embeddings represent the features of the music. Then we can use different methods to aggregate the features as the input of downstream models. The entire process is illustrated in the Figure \ref{fig:model-pipe}. 

\begin{equation}
    \boldsymbol{h_0} = Mel(\boldsymbol{x})
    \label{eq:1}
\end{equation}

\vspace*{-0.2cm}

\begin{equation}
    \boldsymbol{h_{i}} = Encoder_{i}(h_{i-1}) + \boldsymbol{h_{i-1}} \quad i = 1,2,...,L
    \label{eq:2}
\end{equation}

In the evaluating procedure, to demonstrate the representational ability of the extracted features, we employ a simple probing model to perform downstream tasks, which is consistent with prior work \cite{Castellon_Donahue_Liang_2021}. This probing model is structured as a shallow neural network with a single 512-dimensional hidden layer and an output layer. The model's parameters are frozen during the downstream training. 

\subsection{Feature extractor Details}

We adopted the following settings for our testing approach:

\textbf{Baseline feature} We concatenate FBANK and Chroma representation as the baseline for evaluating due to their wide adoption in audio and music related tasks. FBANK features represent the energy of different frequency bands. We use 80-dimensional Mel filter bank as FBANK features, with a frame length of 25ms and an overlap of 10ms; Chroma features represent the energy distribution of different pitch classes (like the 12 notes in Western music) regardless of the octave. Chroma features are particularly useful to analyze harmonic content.

\textbf{MusicFM} We use the checkpoint pre-trained on the MSD dataset \cite{bertin2011msd} with 330M parameters, which is publicly available. The embeddings' hidden dim $d = 1024$,  token sample rate is 25K Hz.

\textbf{MuQ} The MuQ model consists of 12 layers of Conformer, with 310M parameters totally. Mel-RVQ takes the Mel-spectrogram ($d = 128$) of the audio as input and outputs the tokens to be predicted. The embeddings' hidden dim $d = 1024$, token sample rate is 25K Hz. 

\subsection{Canonical Correlation Analysis} \textbf{Canonical Correlation Analysis (CCA)} \cite{HOTELLING_CCA} is a statistical technique for relating two sets of observations arising from an underlying process. CCA has been used as a method to compare representations within and across neural network models  \cite{Raghu_Gilmer_Yosinski_Sohl-Dickstein_2017_SVCCA, Sussillo_Churchland_Kaufman_Shenoy_2015}. 
Specifically, we use the variant PWCCA\cite{Morcos_Raghu_Bengio_2018_pwcca}, which is a improved CCA method that computes a weighted mean of the canonical correlations. PWCCA has been shown to perform robustly in the presence of noise and spurious correlations. We use CCA to measure the similarity between embeddings  produced by encoders at each layer with inputs. This allows us to observe how information evolves during forward propagation. We used a subset of MIR tasks as the dataset for this analysis, which were not seen in the models' pre-training data.




\subsection{Downstream Tasks and Datsets}

We roughly categorize the downstream tasks into the following types:

\begin{itemize}[leftmargin=*]
    \item \textbf{Acoustic tasks}, whose annotations are relatively objective, such as singer, instrument, pitch, and so on.
    \item \textbf{Semantic tasks}, emphasizes semantic features, which are more subjective, including emotion and music genre.
    \item \textbf{Comprehensive tasks}, which have both semantic and acoustic labels, such as music tagging.
\end{itemize}

We evaluate the model by the MARBLE benchmark \cite{yuan2023marble} on downstream tasks. We provide a brief overview of the downstream tasks and datasets used here.

For \textbf{Acoustic-level tasks}, we have :
\begin{itemize}[leftmargin=*]
    \item \textbf{Pitch classification}  classifies one of the 128-pitch of given music clip on the NSynth dataset \cite{Engel_2017_NSynth}. We use accuracy as metric.
    
    \item \textbf{Singer identification} is to recognize the corresponding singer based on the given singing audio on the VocalSet \cite{Wilkins_2018_vocalset}. The evaluation metric is accuracy.


    \item \textbf{Instrument classification} identify which instruments are used in a given music segment on the Nsynth dataset \cite{Engel_2017_NSynth}. The evaluation metric is accuracy.
\end{itemize}

For \textbf{Semantic-level tasks}, we have :

\begin{itemize}[leftmargin=*]
    \item \textbf{Genre classification} aims to determine the most suitable genre for a song. We use the GTZAN dataset \cite{Tzanetakis_Cook_2002} and report the accuracy scores.
    
    \item \textbf{Emotional analysis} is taken on the Emomusic dataset \cite{Soleymani_Caro_Schmidt_Sha_Yang_2013}.  The metrics are determination coefficients for valence ($\mathrm{R2^V}$) and arousal ($\mathrm{R2^A}$) score.

    \item \textbf{Vocal technique detection} identify the techiques that the singer used in a given audio segement on the VocalSet \cite{Wilkins_2018_vocalset}. The evaluation metric is accuracy.

     \item \textbf{Music structure analysis} predicts frame-level functional label for music segments on the Harmonix dataset \cite{nieto2019harmonix}. The evaluation metric is accuracy.

\end{itemize}

For \textbf{Comprehensive tasks}, we have :

\begin{itemize}[leftmargin=*]
    \item \textbf{Music tagging} involves multi-label classification of 50 tags on the MagnaTagATune \cite{law2009mtt} dataset. The metrics are ROC-AUC and average precision (AP).

    \item \textbf{Key detection} estimates the tonal scale and dominant pitch level of each given song, including major and minor scales for all pitch classes, which is 24 classes in total. We use Giantsteps dataset \cite{Knees_2015_GS} and report the refined accuracy score by \cite{raffel2014mir_eval}.

\end{itemize}

\begin{figure*}[t]
    \centering
    \vspace*{-0.5cm}
    \hspace{-20mm} \begin{subfigure}[b]{0.5\textwidth}
        \centering
        \includegraphics[width=\textwidth]{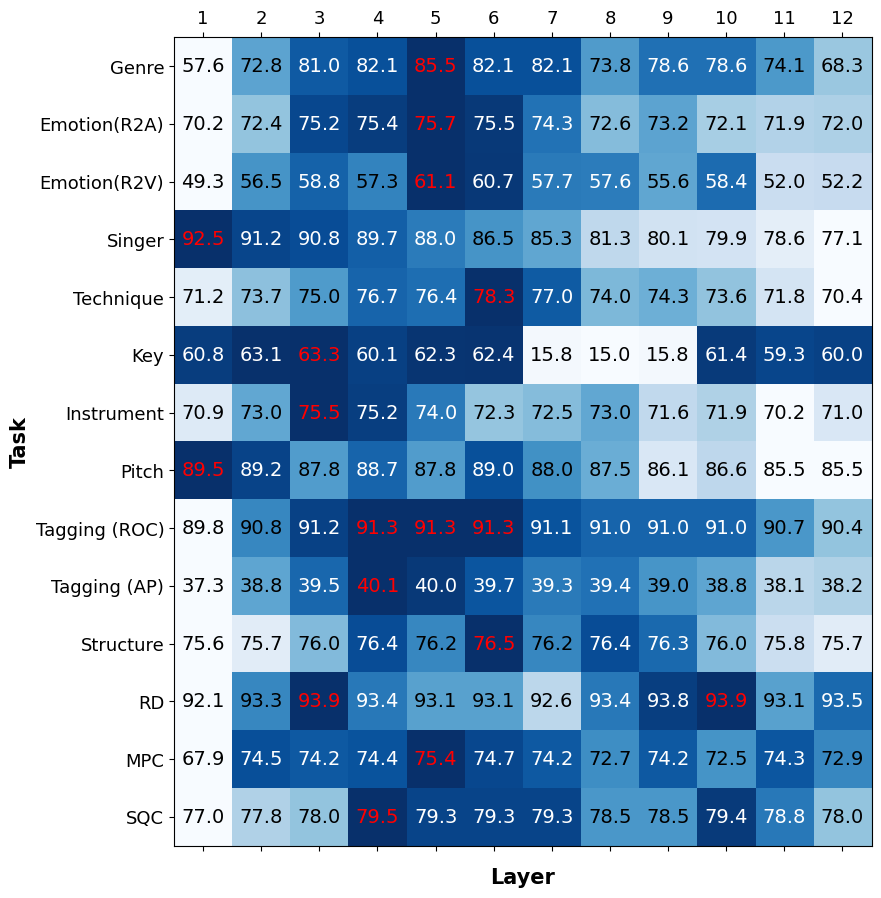} 
        \caption{Layer-wise results on MusicFM}
        \label{fig:layer-musicfm}
    \end{subfigure}
    \begin{subfigure}[b]{0.5\textwidth}
        \centering
        \includegraphics[width=\textwidth]{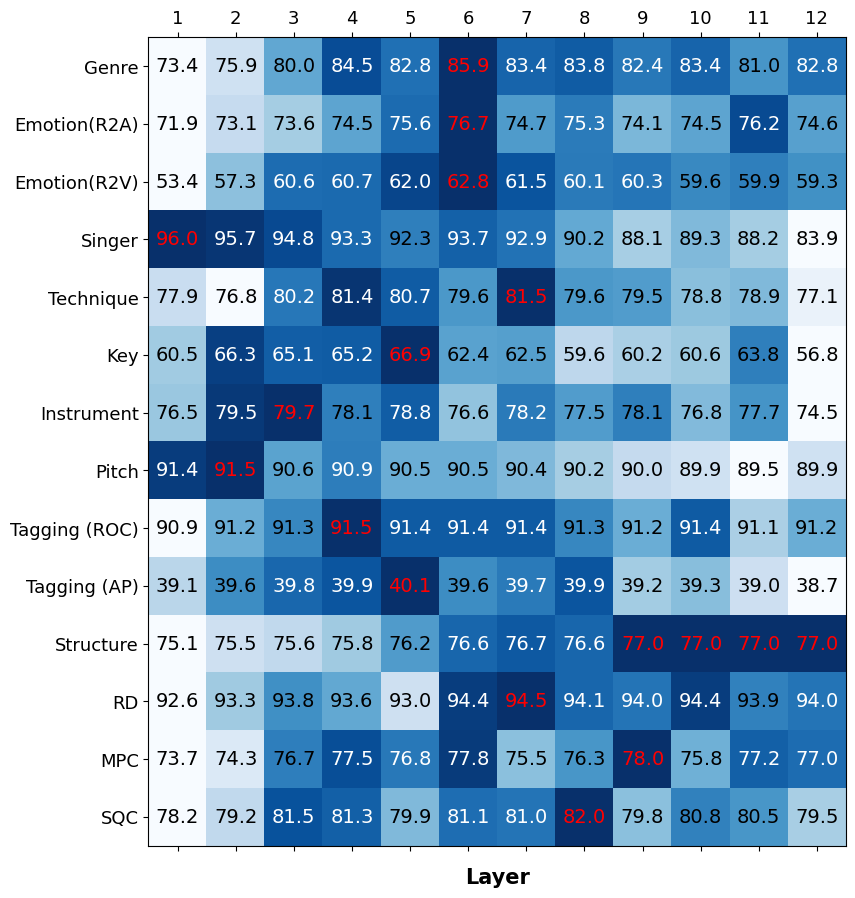} 
        \caption{Layer-wise results on MuQ}
        \label{fig:layer-muq} 
    \end{subfigure} \hspace{-20mm}
    \caption{Layer-wise results on MusicFM and MuQ model, the red numbers indicate the best-performing layers in this column for each task.}
    \label{fig:layer-results}
\end{figure*}

In addition to these tasks, we define our own semantic-level tasks. We utilized our self-collected \textbf{MT} (Music Taste) dataset for these tasks, which contains 4,000 music samples which are manually annotated. Each sample has three type of different semantic labels. The dataset was then divided into training, validation, and test sets in a 3:1:1 ratio. The label of the datasets and tasks are described as follows:

\begin{itemize}[leftmargin=*]
    
    \item \textbf{Music Preference Classification (MPC)} distinguishes whether the given music is perceived as good or bad based on human standards. The labels were obtained through crowdsourcing, where participants were asked to rate each piece of music on the
    following scale:
    
    \begin{enumerate}[label=\alph*.]
    \item 0 points: Unpleasant, out of tune, offbeat, and very poor sound quality.
    \item 1 point: Mostly pleasant, with a few segments that are not enjoyable.
    \item 2 points: Professional level.
    \end{enumerate}
    
    Then, we calculated the average score given by all participants for each track. The tracks with scores greater than 1 were classified as
    "preferred" while scores of 1 or lower were classified as "non-
    preferred."

     \item \textbf{Rap Detection (RD)}  is a three-class classification task that determines which of the following categories a song belongs to. The participants are asked to distinguish:
     \begin{enumerate}[label=\alph*.]
    \item 0 points: Entirely rap.
    \item 1 point: A mix of rap and singing.
    \item 2 points: No rap content.
    \end{enumerate}
    Similar to the above, the track of average scores less than 0.67 were classified as "rap", scores greater than 1.34 were classified as "non-rap", the intermediate scores are classified as "mix".

    \item \textbf{Sound Quality Classification(SQC)}  distinguishes whether the audio quality of the given recording is good. The participants are asked to score:
    \begin{enumerate}[label=\alph*.]
    \item 0 points: Bad quality.
    \item 1 point: Good quality.
    \end{enumerate}
    Similar to the above, scores greater than 1 were classified as
    "Good quality" while scores of 1 or lower were classified as "Bad quality".
\end{itemize}

We then trained the downstream models on these tasks and obtained the final test results. Since the upstream model outputs multi-layered results, we employ two methods to obtain inputs for the downstream models: (1) extracting intermediate features through layer-wise scanning; (2) using learnable parameters $\boldsymbol{c}$ as weights for each layer \cite{Peters_Neumann_Iyyer_Gardner_Clark_Lee_Zettlemoyer_2018_weightsum}, and the weighted sum $\tilde{\boldsymbol{h}}_t$ of the features from all layers serves as the input  :

\begin{equation}
\boldsymbol{c} = c_{1}, ... c_{L}
\end{equation}
\begin{equation}
    \tilde{\boldsymbol{h}}_t=\sum_{l=1}^L \sigma(\boldsymbol{c})_{l} \cdot\boldsymbol{h}_t^l
\end{equation}


Here, $L$ denotes the total number of layers in the model, and $l$ denotes a specific layer. $\sigma(\boldsymbol{c})_{l}$ denotes the softmax score of layer $l$.  We use the softmax function to ensure that the sum of the weights equals 1.

\section{Findings}
\label{sec:findings}

\subsection{ SSL outperforms low-level features. }

\begin{table*}[!ht]
    \centering
    \begin{tabular}{ccccccc}
    \hline
        \textbf{Task} & \textbf{Metrics} & \textbf{FBANK + Chroma} & \textbf{MusicFM} & \textbf{MusicFM-weight} & \textbf{MuQ} & \textbf{MuQ-weight} \\ \hline
        \textbf{Genre} & Acc $\uparrow$ & 15.2 & 85.5 (5) & 80.0 (9) & \textbf{85.9} (6) & 82.8 (9) \\ 
        \textbf{Emotion} & Valence\_r2 $\uparrow$ & 36.7 & 75.7 (5)  & 76.2 (4) & 76.7 (6) & \textbf{77.6} (4) \\ 
        \textbf{Emotion} & Arousal\_r2 $\uparrow$ & 2.9 & 61.1 (5) & 57.2 (4) & 62.8 (6) & \textbf{63.0} (4) \\ 
        \textbf{Key } & Refined\_Acc $\uparrow$ & 15.0 & 63.3 (3) & 63.0 (2) & \textbf{66.9} (5) & 63.7 (2) \\ 
        \textbf{Singer} & Acc $\uparrow$ & 63.5 & 92.5 (1) & 91.3 (1) & \textbf{96.0} (1) & 94.6 (1) \\ 
        \textbf{Technique} & Acc $\uparrow$ & 49.3 & 78.3 (6) & 76.0 (9) & \textbf{81.5} (7) & 76.4 (9) \\ 
        \textbf{Instrument} & Acc $\uparrow$ & 38.1 & 75.5 (3)  & 75.8 (3) & \textbf{79.7} (3) & 77.9 (11) \\ 
        \textbf{Pitch} & Acc $\uparrow$ & 74.4 & 89.5 (1) & 90.2 (4) & \textbf{91.5} (2) & 91.2 (4) \\ 
        \textbf{MTT} & ROC $\uparrow$ & 81.3 & 91.3 (3) & 91.3 (2) & \textbf{91.5} (4) & \textbf{91.5} (2) \\ 
        \textbf{MTT} & AP $\uparrow$ & 23.2 & 40.0 (5) & 39.8 (2) & \textbf{40.1} (5) & 40.0 (2) \\ 
        \textbf{MPC} & Acc $\uparrow$ & 62.2 & 75.4 (5) & 74.3 (4) & \textbf{78.0} (9) & 75.2 (4) \\ 
        \textbf{RD } & Acc $\uparrow$ & 90.5 & 93.9 (3) & 93.8 (5) & \textbf{94.5} (7) & 93.9 (5) \\ 
        \textbf{SQC } & Acc $\uparrow$  & 76.5 & 79.5 (4) & 78.4  (3) & \textbf{82.0} (8) & 81.0 (4) \\ 
        \textbf{Structure} & Acc $\uparrow$ & 75.6 & 76.5 (6) & 76.4 (3) & 77.0 (9) & \textbf{77.2} (9)\\ \hline
    \end{tabular}
    \caption{Overall results. For the layer scanning method, the number in parentheses after the score indicates the layer where the best result occurred. For the method using trainable weighted sums, the number in parentheses indicates the layer with the highest learned weight.}
    \label{tab:eval-result}
    
\end{table*}

According to Table \ref{tab:eval-result}, MusicFM model and MuQ model outperform base features on all of the tasks. In most tasks, the performance of basic features is significantly inferior to that of SSL models. Notably, in the structural analysis task, the basic features achieved comparable performance, likely because this task involves frame-level classification. The sample rate of basic features are equal to the labels, while we need to resample the features of SSL models to match the sample rate.

\subsection{How Feature representations evolve}


\begin{figure}
    \centering
    \includegraphics[width=0.9\linewidth]{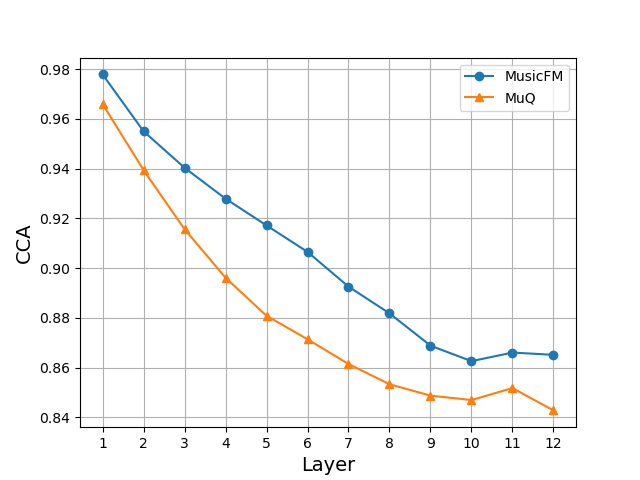}
    \caption{PWCCA scores between the intermediate layer representations and the model input representations}
    \label{fig:CCA}
\end{figure}


We can observe from the layer-wise testing results of MusicFM in Figure \ref{fig:layer-results}(a) that certain tasks, such as singer detection and instrument classification, perform better in the lower layers. These tasks have the characteristic of having objective labels. In contrast, tasks such as structure analysis, singing technique, and genre classification, which have more subjective labels, perform better in the higher layers. Tasks like music tagging, which involve both subjective and objective labels, show a more evenly distributed performance across the middle layers. This reflects a trend where the information in the features gradually shifts from low-level (objective) to high-level (semantic) representations as it is passed forward from the input (Mel Spectrum) to the final layer.

According to Figure \ref{fig:layer-results} (b), MuQ model exhibited a similar distribution pattern to MusicFM, indicating that this is a common characteristic of this type of model. This trend is also evident in Figure \ref{fig:CCA}: from lower to higher layers in both models, the similarity between the output and the input from the first layer decreases progressively.
The CCA result indicate that, in the case of model MusicFM and MuQ, the trend of correlation with acoustic features observed during forward propagation differs from the "high-low-high" pattern seen in autoencoder-style models, which has been found in previous research \cite{Pasad_Shi_Livescu_2022}. The trend in MusicFM and MuQ is consistent with our experimental results in both acoustic and semantic tasks.

Additionally, in most of our test results, the optimal layer did not appear in the final layer. MusicFM's best performance typically occurred within the first 6 layers, while MuQ's best performance appeared within the first 9 layers.

\subsection{Layer scanning vs All layers }


In this section, we focus on how to better select features that are most suitable for specific tasks. Here, we employ both layer-wise scanning and weighted sum methods for feature selection.

As shown in the Table \ref{tab:eval-result}, we observed that for most tasks, directly selecting a single layer produced better results than using all layers. Exceptions were observed with the NSynth dataset on the MusicFM model and the EMO dataset on both models. This may be due to the fact that when selecting all features, the input becomes too large relative to the available data, making it difficult for the downstream model to effectively model all features. However, in the MTT and NSynth datasets, which have a larger amount of data, this issue is mitigated. We can also find that the layers with the highest learned weights are generally not the ones that perform best in single-layer evaluations. Thus, in this situation, directly selecting layers may offer important  prior knowledge for downstream tasks.


\section{Conclusion}

We evaluate music SSL models, including MusicFM and MuQ, across 14 tasks, showing their comprehensive advantages. Layer-wise analysis provides us with a detailed perspective on the relationship between the performance of each specific layer and each specific task. It also reveals a transition from acoustic to semantic features. Future work will explore model architectures, pre-training objectives in music SSL models, and their effects on performance.

\vspace{12pt}

\bibliographystyle{unsrt}
\bibliography{refs}

\end{document}